# Comparative Analysis of Current Component in InGaN-based Blue and AlGaInP-based Red Light-emitting Diode

Dong-Pyo Han[1, 2 a)], Jong-In Shim[2], and Dong-Soo Shin[3]

[1]*Faculty of Science and Technology, Meijo University, 1-501 Shiogamaguchi, Tempaku-ku, Nagoya 468-8502, Japan*

[2]*Dept. of Electronics and Communication Engineering, Hanyang University, ERICA Campus, Ansan, Gyeonggi-do 426-791, Korea*

[3]*Dept. of Applied Physics and Dept. of Bionanotechnology, Hanyang University, ERICA Campus, Ansan, Gyeonggi-do 426-791, Korea*

**Abstract**

In this paper, we aim to understand the apparent characteristics of IQE and *I-V* curve in AlGaInP and InGaN LED devices. For the analysis, we separate the current into radiative current ($I_R$) and non-radiative current ($I_{NR}$) component by using the information of IQE. We carefully analyze each current component by ideality factor, *S* parameter, and the modified Shockley diode equation which is suitable for LED device. Through the analyses, it has been found that the characteristics of respective current components are basically similar for both samples while the physical origin of the potential drop induced by $I_R$ and the amount of double injection current induced $I_{NR}$ by are different. Compared with AlGaInP LEDs, the InGaN LEDs have higher degrees of electron overflow initiated by low recombination rate in active region, causing both the efficiency droop and the higher operating voltage. To remedy this, the radiative recombination rate and/or the active volume should be increased further.

[a)] E-mail: han@meijo-u.ac.jp

The wide-bandgap *pn* junction is one of the most promising device structures in semiconductors since electronic and optoelectronic devices can utilize the wide-bandgap *pn* junction for various applications.[1,2] Examples of wide-bandgap *pn*-junction semiconductor devices include photodiodes, solar cells, laser diodes, and, most notably, light-emitting diodes (LEDs).[3] Recently, there has been significant progress in the research and development of high efficient wide-bandgap LEDs covering the visible and ultraviolet spectral regions. Owing to the progress, general lighting based on LEDs has become a reality with the advantages of brightness, energy efficiency, and environment friendliness.[3,4] The LEDs devices based on the InGaN material system have a very wide spectral range from near-ultraviolet to green as the emission wavelength can be varied by adjusting the indium composition in the active multiple quantum wells (MQWs).[4] On the other hand, AlGaInP-based LEDs have been the dominant LED devices for high-brightness emitter in the visible spectral range from red to yellow since they were first demonstrated by Kobayashi *et al*. in 1985.[4,5] From these efforts, three primary colors of light, red, green, and blue, could be fully realized by LED devices.

Although the performances of LED devices have significantly improved so far, higher light output power and lower electrical power consumption are still required for utilizing general lighting source. In particular, unlike the AlGaInP material system, the internal quantum efficiency (IQE) of InGaN-based LEDs is limited by the so-called "efficiency droop", the monotonic decrease in IQE with increasing injection current,[6] and the "green-gap", the monotonic decrease in IQE with increasing indium composition in MQWs.[7] Many mechanisms have been proposed to explain the efficiency droop and the green gap, including the piezoelectric field,[8] potential fluctuation in MQWs,[9] Auger recombination,[10] poor hole injection,[11] and phase space filling effect.[12] The high operating voltage in InGaN LEDs is considered as another limiting factor in the aspect of the wall-plug efficiency (WPE) of LED devices.[13] It is generally believed that the series resistance of the LED device causes a large potential drop under high driving current. Both the AlGaInP and InGaN LEDs are fundamentally similar in the aspect of electroluminescence (EL) by recombination of diffused electron-hole pairs in MQWs and in aspect of IQE, even efficiency droop phenomenon is observed in AlGaInP-based LED at cryogenic temperature.[14] However, there also exist noteworthy differences: i) AlGaInP LEDs are grown lattice-matched to the GaAs substrates, whereas InGaN LEDs are grown lattice-mismatched to the sapphire substrates. This lattice mismatch induces high-density threading dislocations (TDs) in InGaN LEDs. ii) EL efficiency of InGaN LEDs is insensitive to TD density compared to other material system. iii) InGaN LEDs are affected significantly by the piezoelectric field due to the strain unlike the AlGaInP LEDs. iv) The 2-D potential distribution in MQWs is inhomogeneous in InGaN blue LED. Finally, v) AlGaInP LEDs employ more numerous and thicker MQWs as the active layer.[4,14]

In this paper, we aim to understand the apparent characteristics of IQE and *I-V* curve in AlGaInP and InGaN LED devices. For the analysis, we separate the current into radiative current ($I_R$) and non-radiative current ($I_{NR}$) component by using the information of IQE. We carefully analyze each current component by ideality factor, *S* parameter, and the modified Shockley diode equation which is suitable for LED device.[15] Based on these comparative analyses, we try to find the similarities and differences between the LED devices as well as understand the cause of the decrease in WPE including the efficiency droop and high operating voltage.

For experiments, we utilized commercial LED samples: a blue LED with InGaN MQWs and a red LED with AlGaInP MQWs. The peak wavelength and the size are 450 nm and 600×600 μm$^2$ for the blue sample and 630 nm and 800×800 μm$^2$ for the red sample. Both samples were fabricated into chips with lateral electrodes and packaged as a surface-mounted-device (SMD) type without epoxy dome. *I-V* curves were measured under pulse-voltage condition. The light output power was collected by a Si photo diode under pulsed-current driving condition (pulse period: 100 μs and duty cycle: 1%). IQE data were obtained by the conventional temperature-dependent electroluminescence (TDEL) measurement, which is widely used in various research works.[16-18]

Figure 1 (a) and (b) depict the IQE values of both samples as a function of driving current density (*J*) plotted on linear and semi-log scales, respectively. The IQE of the blue sample shows a maximum value of ~86% at a low current density of ~2 A/cm$^2$ and rapidly decreases compared to the red sample which has the maximum IQE of ~91% at a higher current density of ~10 A/cm$^2$ and slowly decreases. In other words, the IQE of the blue sample shows faster increase in IQE to its peak value at a lower current density and the efficiency droop is severer beyond the peak value of the IQE than the red sample. These are typical IQE characteristics observed with InGaN and AlGaInP LEDs.[18-20]

In Figure 2 (a) and (b), the *J-V* curves are depicted for blue and red samples plotted on linear and semi-log scales, respectively. It is seen that the blue sample has a relatively higher forward leakage in the low-bias region (1.5 - 2.3 V) and a higher series resistance in the high-bias region (> 3 V) compared to the red sample. It is generally considered for the blue sample that the tunneling mechanism originating from the TDs due to the lattice mismatch induces the leakage current[21] and that the low hole concentration resulting from the high activation energy of acceptors is responsible for the high series resistance.[22]

From the definition of the IQE, the ideality factor $n_R$ and $n_{NR}$ corresponding to $I_R$ and $I_{NR}$ can be obtained by combining the information of IQE and *I-V* as expressed in eqs. (1) and (2):[15,23]

$$n_R = \frac{q}{k_B T}\left(\frac{\partial \ln(\eta_{IQE} \cdot I)}{\partial V}\right)^{-1} \quad (1)$$

$$n_{NR} = \frac{q}{k_B T}\left(\frac{\partial \ln(I \cdot (1-\eta_{IQE}))}{\partial V}\right)^{-1} \quad (2)$$

where $q$ is the elementary charge, $k_B$ is the Boltzmann constant, $T$ is the absolute temperature, and $\eta_{IQE}$ is the IQE. $I_R$ and $I_{NR}$ are defined as $\eta_{IQE} \cdot I$ and $I \cdot (1-\eta_{IQE})$, respectively. In Fig. 3 (a) and (b), $n_R$ and $n_{NR}$ are plotted as a function of current density for blue and red sample, respectively. It is seen that $n_R$ remains at 1 at low current region for both samples and then increases with increasing driving current. On the other hands, $n_{NR}$ remains at 2 at low current region for both samples and then increases with increasing driving current. From the Shockley theory, the ideality factor indicates the mechanism of carrier transport and recombination. Thus, it is obvious that $n_R$ of 1 is a result from the direct band-to-band recombination in MQWs of diffusion current since $I_R$ is proportional to light output power (*L*) and peak wavelength of EL is identical to bandgap energy of MQWs. The $n_{NR}$ of 2 is a result from the Shockley-Read-Hole (SRH) recombination via defects for $I_{NR}$ since it is resulted from non-radiative recombination in low current density region.[22] The increase in $n_R$ and $n_{NR}$ with increasing driving current is considered resulting from the potential drop of the LED device. Therefore, for the blue and red samples under investigation, we confirm that the carrier recombination and transport mechanisms of $I_R$ and $I_{NR}$ are fundamentally identical. However, notable difference can be observed that the increase in $n_R$ and $n_{NR}$ with increasing driving current begin with lower current density in blue sample compared to that of red sample. In other word, larger potential drop occur in InGaN-blue LED sample during carrier transport, while transport and recombination mechanism are identical for both samples.

Figure 4 shows parameter called $S$, which is defined as, $\partial \log(I_R)/\partial \log(I_{NR})$ as a function of driving current density.[24] Here, from the rate equation, $I_R$ and $I_{NR}$ under steady state can be expressed as:[25]

$$I_R = qV_{eff}\left(BN^2\right) \quad (3)$$

$$I_{NR} = qV_{eff}\left(AN\right) + f(N) \quad (4)$$

where $A$ is the SRH recombination coefficient, $B$ is the radiative recombination coefficient, $N$ is the carrier concentration in the active layer, $q$ is elementary charge, $V_{eff}$ is effective active volume, and $f(N)$ is carrier leakage term including carrier emitting to vacuum level by Auger recombination, carrier overflow, and carrier spill-over. From the eq. (3) and (4), $S$ should have the value between 1 and 2 if the carriers recombination is occurred predominantly in active region, i.e., $qV_{eff}(AN) \gg f(N)$. On the other hand, if carrier leakage is dominant transport mechanism, i.e., $qV_{eff}(AN) \ll f(N)$, $S$ should be lower than value of 1 since carrier leakge term have higher order than $N^2$, i.e., $f(N) = CN^3 + DN^4 + \cdots$.[25] As shown in Fig. 4, $S$ has the value between 1 and 2 as high as current density of ~1 A/cm$^2$ and ~10 A/cm$^2$ for blue and red sample, respectively. It implies that in red sample, approximately ten times larger amount of carriers recombine in active region compared to blue sample. The value of $S$ exceeding two in low current density region ($< 10^{-2}$ A/cm$^2$) for blue sample results from the leakage current by defect-assisted tunneling introduced by lattice mismatch between sapphire substrate and GaN layer.[21]

For more detailed comparative analysis for two samples, we employ the modified Shockley diode equation for $I_R$ vs. $V$ and $I_{NR}$ vs. $V$ curve fitting expressed as:[15]

$$I_R = I_{S,R} \exp\left[\frac{q(V - IR_S - \ln(1 + \alpha' I_R))}{k_B T}\right], \quad (5)$$

$$I_{NR} = I_{S,NR} \exp\left[\frac{q(V - IR_S - \sqrt{I_{NR}} D_{DI})}{2k_B T}\right], \quad (6)$$

where $I_{S,R}$ ($I_{S,NR}$) is the reverse saturation current corresponding to $I_R$ ($I_{NR}$), $\alpha'$ is parameter corresponded to phase-space filling, $R_S$ is the conventional series resistance and $D_{DI}$ represents the coefficient for the potential drop induced by the double injection current. Figure 5 (a) and (b) shows $I_R$ and $I_{NR}$ data in the blue sample and red sample fitted by eqs. (5) and (6), respectively. It is seen that the modified Shockley equation [eq. (5) and (6)] fits the experimental data of $I_R$ and $I_{NR}$ very well for both sample. Table 1 summarizes the fitting parameters to fit the data in Fig. 5 (a) and (b). As shown in Table 1, we can see that large difference in the fitting value of $\alpha'$ and $D_{DI}$ compared to that of $R_S$. Here, the value of $\alpha'$ represents the saturation of radiative recombination rate due to phase space filling effect under high current injection and the value of $D_{DI}$ represents the carrier overflow as a form of double injection current. Thus, we can conclude that both high operating voltage and server efficiency droop in blue sample are mainly caused by low recombination rate in active region which induce large amount of carrier recombination.

From the above analyses, common characteristics of blue and red samples lead to the following conclusions: i) From $n_R$ of 1, the band-to-band radiative recombination by diffused carrier is mainly responsible for $I_R$ for both samples. ii) $n_{NR}$ of 2 indicates that the SRH nonradiative recombination is the main cause of $I_{NR}$ for both samples. iii) Lastly, the potential drop in $I_{NR}$ results from the double injection current owing to overflown electrons to the p-type layer. On the other hand, there are noteworthy differences: i) The potential drop induced by $I_R$ mainly results from phase space filling in the blue sample, while conventional ohmic resistance in red sample. ii) The carrier recombination rate in active region is much higher in red sample. iii) Lastly, a larger amount of electron overflow to the p-type layer induces in the blue sample.

In summary, we have analyzed $I_R$ and $I_{NR}$ in InGaN and AlGaInP LEDs by ideality factor, $S$ parameter, and the modified Shockley diode equation. Through the analyses, it has been found that the characteristics of respective current components are basically similar for both samples while the physical origin of the potential drop induced by $I_R$ and the amount of double injection current induced $I_{NR}$ by are different. Compared with AlGaInP LEDs, the InGaN LEDs have higher degrees of electron overflow initiated by low recombination rate in active region, causing both the efficiency droop and the higher operating voltage. To remedy this, the radiative recombination rate and/or the active volume should be increased further. We think that the difference characteristics pointed out in these analyses give new insight into the efficiency droop in InGaN-based LED devices.

**Figures Captions**

FIG. 1. IQE characteristics of blue and red LED sample measured as a function of driving current plotted on (a) linear and (b) semi-log scales, respectively.

FIG. 2. I-V characteristics of blue and red sample plotted on (a) linear and (b) semi-log scales, respectively

FIG. 3. Ideality factor corresponded to $I_R$ and $I_{NR}$ as a function of driving current density plotted on semi-log scales for (a) blue and (b) red LED samples.

FIG. 4. Slope of log($I_R$)-log($I_{NR}$) curve (*S* parameter) against the driving current density for blue and red LED samples.

FIG. 5. Experimental $I_R$ and $I_{NR}$ vs. *V* fitted by the modified Shockley diode equation for (a) blue and (b) red LED samples.

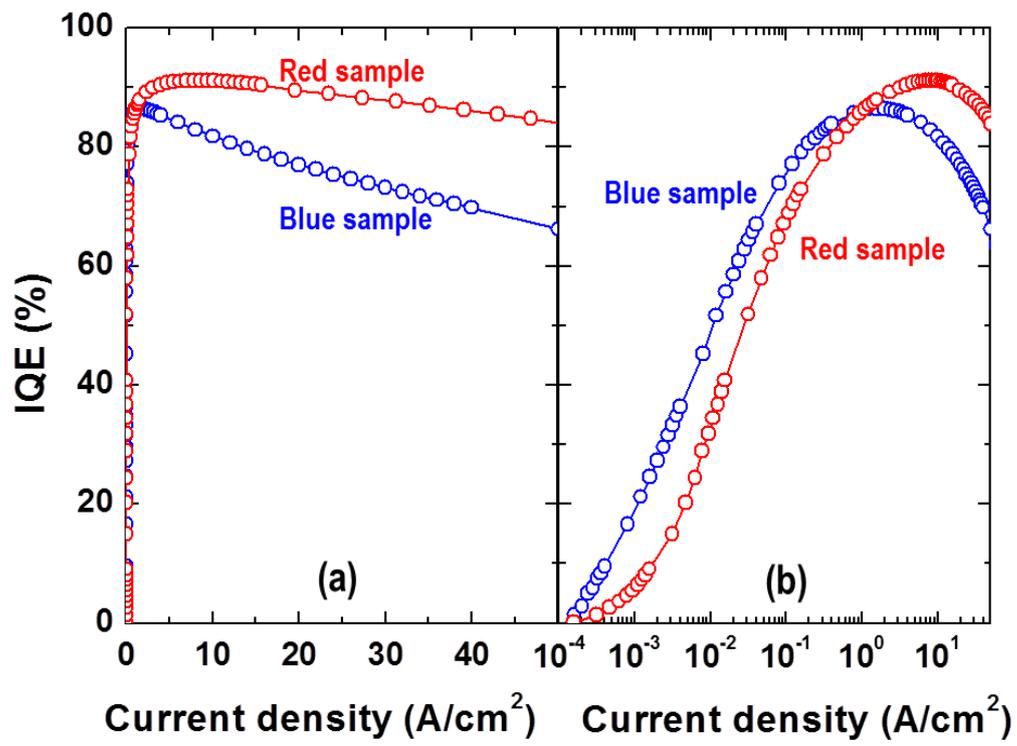

Fig.1

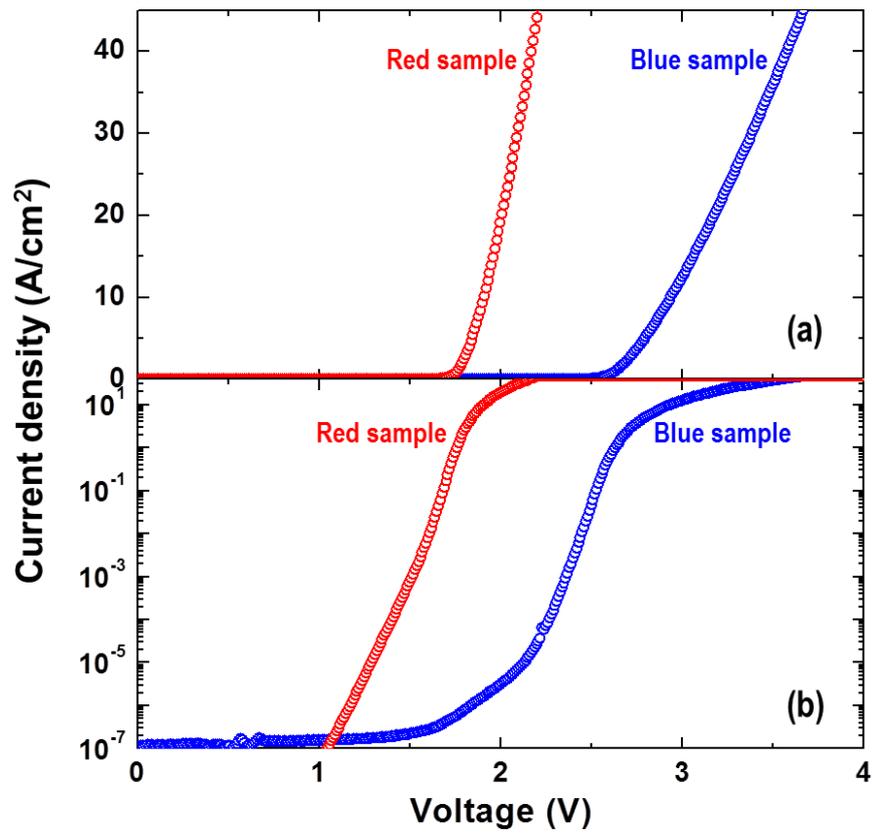

Fig.2

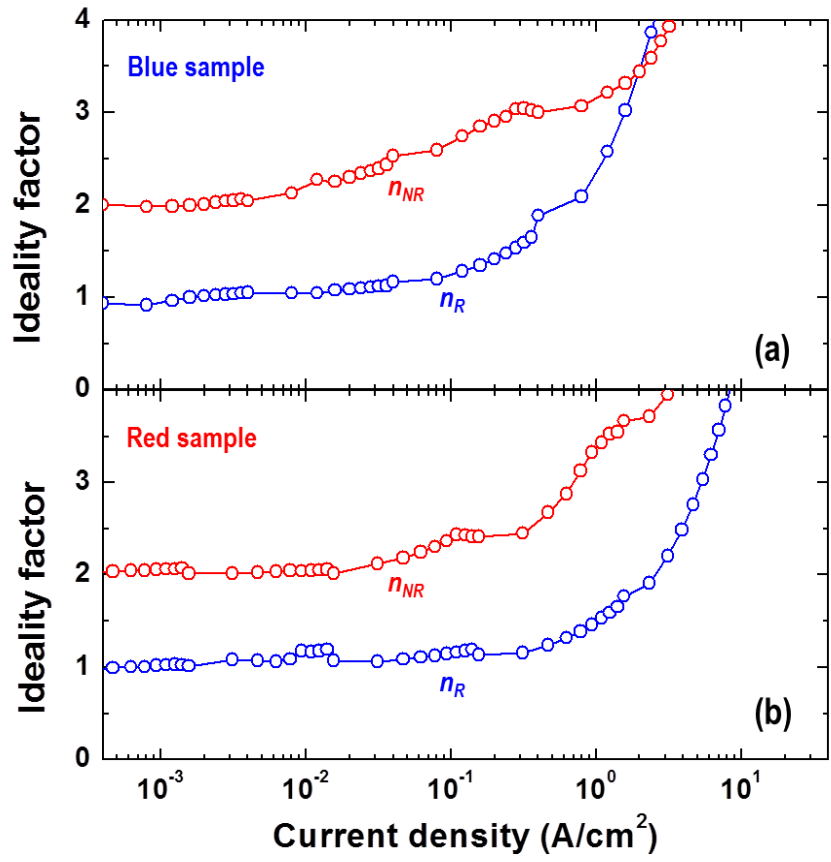

Fig.3

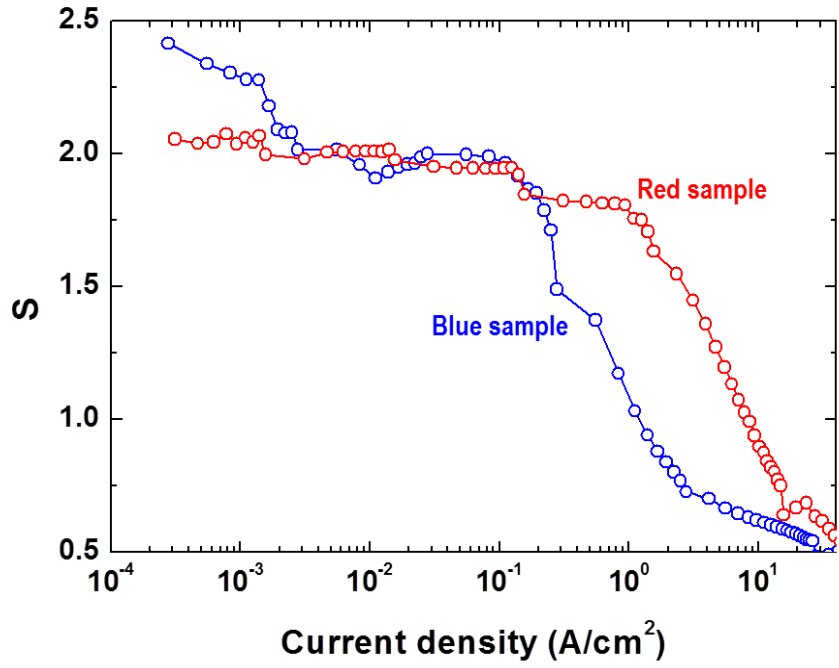

Fig.4

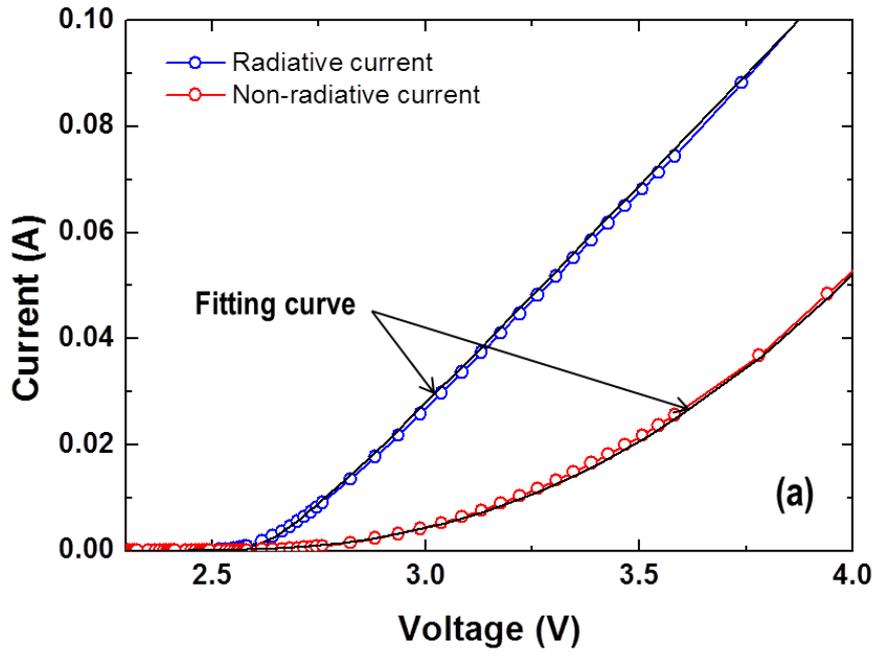

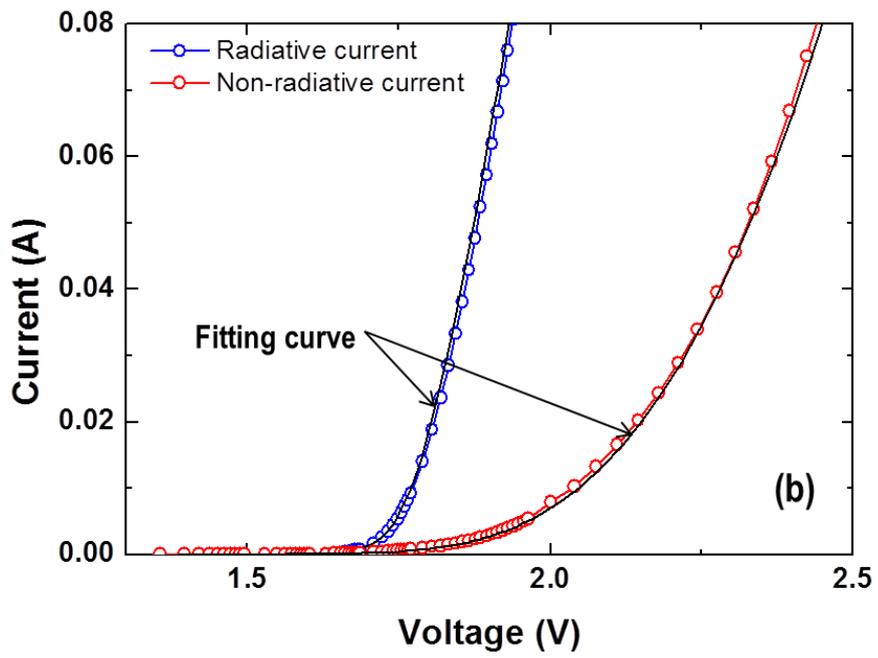

Fig.5

TABLE 1. Summary of fitting parameter in Fig. 5 (a) and (b).

|  | $I_{S,R}$ (A) | $I_{S,NR}$ (A) | $R_S$ (Ω) | $D_{DI}$ (V/A$^{0.5}$) | $\alpha'$ |
|---|---|---|---|---|---|
| Blue sample | $8.2 \times 10^{-47}$ | $1.4 \times 10^{-24}$ | 3.2 | 7.2 | 3.5 |
| Red sample | $2.4 \times 10^{-30}$ | $1.8 \times 10^{-18}$ | 2.1 | 14.5 | 0.2 |